\documentclass[preprint2]{aastex}
\usepackage{amsmath}
\usepackage{graphicx}
\begin{document}
\title{Trapping low-mass planets at the inner edge of the protostellar disc}
\author{R. Brasser\altaffilmark{1}, S. Matsumura\altaffilmark{2}, T. Muto\altaffilmark{3} and S. Ida\altaffilmark{1}}
\altaffiltext{1}{Earth Life Science Institute, Tokyo Institute of Technology, Tokyo, Japan}
\altaffiltext{2}{Division of Physics, University of Dundee, Dundee, UK}
\altaffiltext{3}{Division of Liberal Arts, Kogakuin University, Tokyo, Japan}

\begin{abstract}
The formation of multiple close-in low-mass exoplanets is still a mystery. The challenge is to build a system 
wherein the outermost planet is beyond 0.2~AU from the star. Here we investigate how the prescription for type I planet
migration affects the ability to trap multiple planets in a resonant chain near the inner edge of the protostellar 
disc. A sharp edge modelled as a hyperbolic tangent function coupled with supersonic corrections to the classical type 
I migration torques results in the innermost planets being pushed inside the cavity through resonant interaction with 
farther planets because migration is starward at slightly supersonic eccentricities. Planets below a few Earth masses 
are generally trapped in a resonant chain with the outermost planet near the disc edge, but long-term stability is not 
guaranteed. For more massive planets the migration is so fast that the eccentricity of the innermost resonant pair is 
excited to highly supersonic levels due to decreased damping on the innermost planet as it is pushed inside the cavity; 
collisions frequently occur and the system consists one or two intermediate-mass planets residing closer to the 
star than the disc's inner edge. We found a neat pileup of resonant planets outside the disc edge only if the corotation 
torque does  not rapidly diminish at high eccentricity. We call for detailed studies on planet migration near the 
disc's inner edge, which is still uncertain, and for an improved understanding of eccentricity damping and disc torques 
in the supersonic regime.
\end{abstract}

\section{Introduction}
Preventing low-mass planets from migrating to their host star is a long-standing problem. Mergers with their host star 
can be prevented if the protostellar disc has a sharp inner edge at a few stellar radii \citep{Masset06,O10}. 
\cite{TP07} showed that migrating protoplanets usually end up in resonances. Some of these resided inside the 
disc's inner cavity. \cite{OI9} also build a resonant chain of low-mass protoplanets that stalled near the 
disc's inner edge when the migration was artificially slowed down and the reduction of the corotation torque 
was ignored. Recently \cite{I17} trapped a high number of low-mass planets outside the disc edge in a resonant chain, 
that subsequently needed to break to account for the currently-observed exoplanet period distribution. On the other 
hand, \cite{Mat17} had trouble trapping multiple planets near the disc's inner edge, even though their migration 
prescription was very similar to that of \cite{I17}: both include supersonic corrections to the migration and 
eccentricity damping timescales. \cite{TP07} also included such corrections, but they followed the prescription of 
\cite{PL00} while \cite{I17} and \cite{Mat17} followed \cite{CN14}. The simulations of \cite{Mat17} usually resulted in 
one or two hot Neptune planets rather than a multiplet of smaller planets. The disparity between all of these results 
warrants further study.

\section{Disc model and planet migration}
We employ the disc model of \cite{ida16}, which is based on \cite{GL07} and \cite{Oka11}. Here we briefly summarise 
their model.\\

\subsection{Disc parameters}
We assume steady-state accretion onto the Sun. The gas accretion rate is
\begin{equation}
\dot{M}_*= 3\pi \alpha \Sigma H^2 \Omega_{\rm K},
\label{eq:dotmstar}
\end{equation}
where $\Sigma$ is the gas surface density, $H$ is the disc scale height and $\Omega_{\rm K}$ is the orbital frequency. 
The $\alpha$-viscosity is assumed to be constant \citep{SS73}. The disc scale height is $H=c_s/\Omega_{\rm K}$, where 
$c_s=(\gamma k_BT/\mu m_{\rm prot})^{1/2}$ with $\gamma=7/5$, $k_B$ is the Boltzmann constant, $m_{\rm prot}$ the 
proton mass and $\mu=2.3$ is the mean atomic mass of the gas. Now $\dot{M}_*$ evolves as \citep{H98}
\begin{equation}
 \log \Bigl(\frac{\dot{M}_*}{M_\odot\,{\rm yr}^{-1}}\Bigr) = -8-\frac{7}{5}\log\Bigl(\frac{t}{1\,{\rm 
Myr}}+0.1\Bigr).
\end{equation}
The extra 0.1~Myr avoids the logarithmic singularity \citep{Bitsch15}.\\

For solar-type stars, the midplane temperature in the viscous part of the disc is
\begin{equation}
T = 200 \alpha_3^{-1/5} \dot{M}_{*8}^{2/5}\left(\frac{r}{1\,{\rm AU}}\right)^{-9/10}
\label{eq:Temp}
\end{equation}
where $r$ is the distance to the star and we defined $\dot{M}_{*8} = \dot{M}_*/10^{-8}\,M_\odot\,{\rm yr}^{-1}$ 
and $\alpha_3 = \alpha/10^{-3}$. The reduced scale height $h=H/r$ is
\begin{equation}
h = 0.034 \alpha_3^{-1/10}\dot{M}_{*8}^{1/5} \Bigl(\frac{r}{1\,{\rm AU}}\Bigr)^{1/20}.
\label{eq:hscale}
\end{equation}
Equations (\ref{eq:dotmstar}), (\ref{eq:Temp}) and (\ref{eq:hscale}) may be combined to compute the surface 
density of the gas.

\subsection{Disc inner edge implementation}
Near the star the surface density of the disc is assumed to smoothly decrease to zero. \cite{C14} suggested that
\begin{equation}
 \Sigma = \Sigma(r_{\rm tr})\tanh\Bigl(\frac{r-r_{\rm in}}{H}\Bigr)
\end{equation}
where $r_{\rm tr}$ is the planet trap location from the star; here the surface density is maximal. The trap is at 
$r_{\rm tr}=0.1$~AU for solar-type stars; it is unclear how reliable the employed disc model is closer to the star 
where MHD effects become important. Trapping planets requires a sharp edge \citep{Masset06,O10}. Therefore we set the 
inner edge of the disc at $r_{\rm in}=0.95r_{\rm tr}$, which is approximately 2 scale heights inside of $r_{\rm 
tr}$. The surface density slope is computed as
\begin{equation}
 s \equiv - \frac{d \ln \Sigma}{d \ln r} =-\frac{1}{h}\Bigl(\frac{1}{x}-x\Bigr).
\end{equation}
where $x=\tanh(\frac{r_{\rm tr}-r_{\rm in}}{H})$. Now $s \rightarrow -\infty$ as $r \rightarrow r_{\rm in}$ and all 
planets in the type I regime cease migrating at or near $r_{\rm tr}$.\\

To avoid the divergence of $s$ at $r_{\rm in}$, which could possibly cause numerical artefacts, we also tested a linear 
decrease in $\Sigma$, for which $s=-1$. The resulting discontinuity of $s$ at $r_{\rm tr}$ was made smooth via a linear 
connection over length $0.2H$.

\subsection{Planet migration}
The gas disc exerts torques and tidal forces on the embedded planets which result in a combined effect of radial 
migration and the damping of the eccentricity and inclination. For low-mass planets the migration is of type I 
\citep{tanaka02} while massive planets that are able to clear the gas in their vicinity experience type II migration 
\citep{lp86}. Here we are only interested in the former. \\

We follow \cite{CN14} for computing the torque and the direction of migration. Their formulae are based on \cite{P11} 
for the torque and on \cite{FN14} and \cite{cn08} for the eccentricity damping, including corrections to the 
damping timescale and corotation torque in the supersonic regime when the eccentricity $e >h$. We restrict ourselves to 
planar orbits. The normalised torque is \citep{P11}
\begin{equation}
 \frac{\gamma\Gamma}{\Gamma_0} = \frac{\Gamma_{\rm C}}{\Gamma_0}F_C + \frac{\Gamma_{\rm L}}{\Gamma_0}F_L
 \label{eq:torque}
\end{equation}
where $\Gamma_{\rm C}=\Gamma_{\rm C,baro} + \Gamma_{\rm C,ent}$ and  $\Gamma_{\rm L}$ are the corotation and 
Lindblad torques respectively and $\Gamma_0= (m_p/m_\odot)^2(H/r)^{-2}\Sigma\Omega_{\rm K}^2$ is a normalisation 
constant. The corotation torque in a non-isothermal disc with thermal diffusion becomes 
\begin{eqnarray}
 \label{eq:entropypart}
\Gamma_{\rm C,baro} &=&  F(p_\nu)G(p_\nu)\Gamma_{\rm hs,baro} \nonumber \\
 &+&[1-K(p_\nu)]\Gamma_{\rm C,lin,baro}, \\
 \Gamma_{\rm C,ent} &=&  F(p_\nu)F(p_\chi)[G(p_\nu)G(p_\chi)]^{1/2}\Gamma_{\rm hs,ent}  \nonumber \\
 &+&\{[1-K(p_\nu)][1-K(p_\chi)]\}^{1/2}\Gamma_{\rm C,lin,ent}.\nonumber
\end{eqnarray}
Here $p_\nu$ and $p_\chi$ depend on $m_p$, $h$ and $\alpha$ \citep{P11}. The functions $F(p)$, $G(p)$ and 
$K(p)$ determine the amount of torque saturation and are dependent on the planet mass and disc scale height 
\citep{P11}. In steady state $q+s=\frac{3}{2}$, where $q=-\frac{d\ln T}{d\ln r}$. The remaining contributions are then 
\citep{P11}

\begin{eqnarray}
 \frac{\gamma\Gamma_{\rm L}}{\Gamma_0} &=& -2.5 -1.7q +0.1s = -2.35 -1.8q \nonumber \\
 \frac{\gamma\Gamma_{\rm hs,baro}}{\Gamma_0} &=& 1.1(3/2-s) = 1.1q \nonumber \\
 \frac{\gamma\Gamma_{\rm C,lin,baro}}{\Gamma_0} &=& 0.7(3/2-s) = 0.7q \nonumber \\
 \frac{\gamma\Gamma_{\rm hs,ent}}{\Gamma_0} &=& \frac{7.9\xi}{\gamma} = 
5.6\Bigl(\frac{7}{5}q-\frac{3}{5}\Bigr)\nonumber \\
 \frac{\gamma\Gamma_{\rm C,lin,ent}}{\Gamma_0} &=& \Bigl(2.2-\frac{1.4}{\gamma}\Bigr)\xi = 
0.8\Bigl(\frac{7}{5}q-\frac{3}{5}\Bigr),
 \label{eq:gammacomps}
\end{eqnarray}
where $\xi=-\frac{d\ln S}{d\ln r}=q-(\gamma-1)s=\frac{7}{5}q-\frac{3}{5}$ is the negative entropy gradient. The 
$F$, $G$ and$K$ functions disallow writing the explicit dependence of $\Gamma$ on $q$, but generally $\Gamma_C \propto 
q$ and $\Gamma_L \propto -q$. \\

The factors $F_L$ and $F_C$ are
\begin{eqnarray}
\ln F_C &=& -\frac{e}{e_{\rm f}}, \nonumber \\
F_L &=& \frac{1+(0.444\hat{e})^{1/2}+(0.352\hat{e})^6}{1-(0.495\hat{e})^4},
\label{eq:fcfl}
\end{eqnarray}
where $e_{\rm f}= 0.01+{\textstyle \frac{1}{2}}h$ and $\hat{e}=e/h$. The eccentricity damping timescale 
$\tau_e=-e/\dot{e}$ is \citep{cn08}
\begin{equation}
\tau_{e} = 1.282t_{\rm wav}(1-0.14\hat{e}^2+0.06\hat{e}^3),
\label{eq:taue}
\end{equation}
where the wave timescale is \citep{TW04}
\begin{equation}
t_{\rm wav} =\Bigl(\frac{M_*}{m_p}\Bigr)\Bigl(\frac{M_*}{\Sigma r^2}\Bigr)h^4 \Omega_K^{-1}.
\end{equation}
The 'migration timescale' \citep{cn08} is $\tau_m = -L/\dot{L}$, and is
\begin{equation}
\tau_{m} = -\frac{t_{\rm wav}\Gamma_0}{h^2\Gamma}.
\end{equation}
Despite its name, $\tau_m$ is {\it not} the actual migration timescale. By definition $\tau_m = -L/\dot{L}$ 
\citep{CN07} and $L=m_p\sqrt{GM_\odot a (1-e^2)}$, so that $\tau_m \approx 2\tau_a$ when $e \approx 0$. Here 
$\tau_a = -a/\dot{a}$ is the timescale for semi-major axis evolution. In the supersonic regime hydrodynamical 
simulations show that the torque reverses direction when $e \sim 2h$ and the torque is maximal (and positive) when $e 
\sim 4h$ \citep{cn08}. However, \cite{cn08} show that the planet always migrates inwards, despite the 
torque reversal at high eccentricity. The analytical approach of \cite{Muto11} agrees with this result: inward 
migration persists at high eccentricity. Therefore, it is incorrect to use $\tau_m$ to compute the 
evolution of the semi-major axis; $\tau_a$ should be used instead \citep{Mat17}. This is 
\begin{equation}
 \tau_a^{-1} = 2\tau_m^{-1} + \frac{2e^2}{(1-e^2)}\tau_e^{-1}.
 \label{eq:ta}
\end{equation} 
\cite{I17} adopted $\tau_m$ for the migration timescale, which led to the aforementioned difference. All timescales are 
positive for inward migration and negative for outward migration.

\begin{figure*}[t]
%\resizebox{\hsize}{!}{\includegraphics[angle=0]{timescales.png}}
\centering
\begin{minipage}{0.5\textwidth}
\centering
\includegraphics[width=0.99\textwidth]{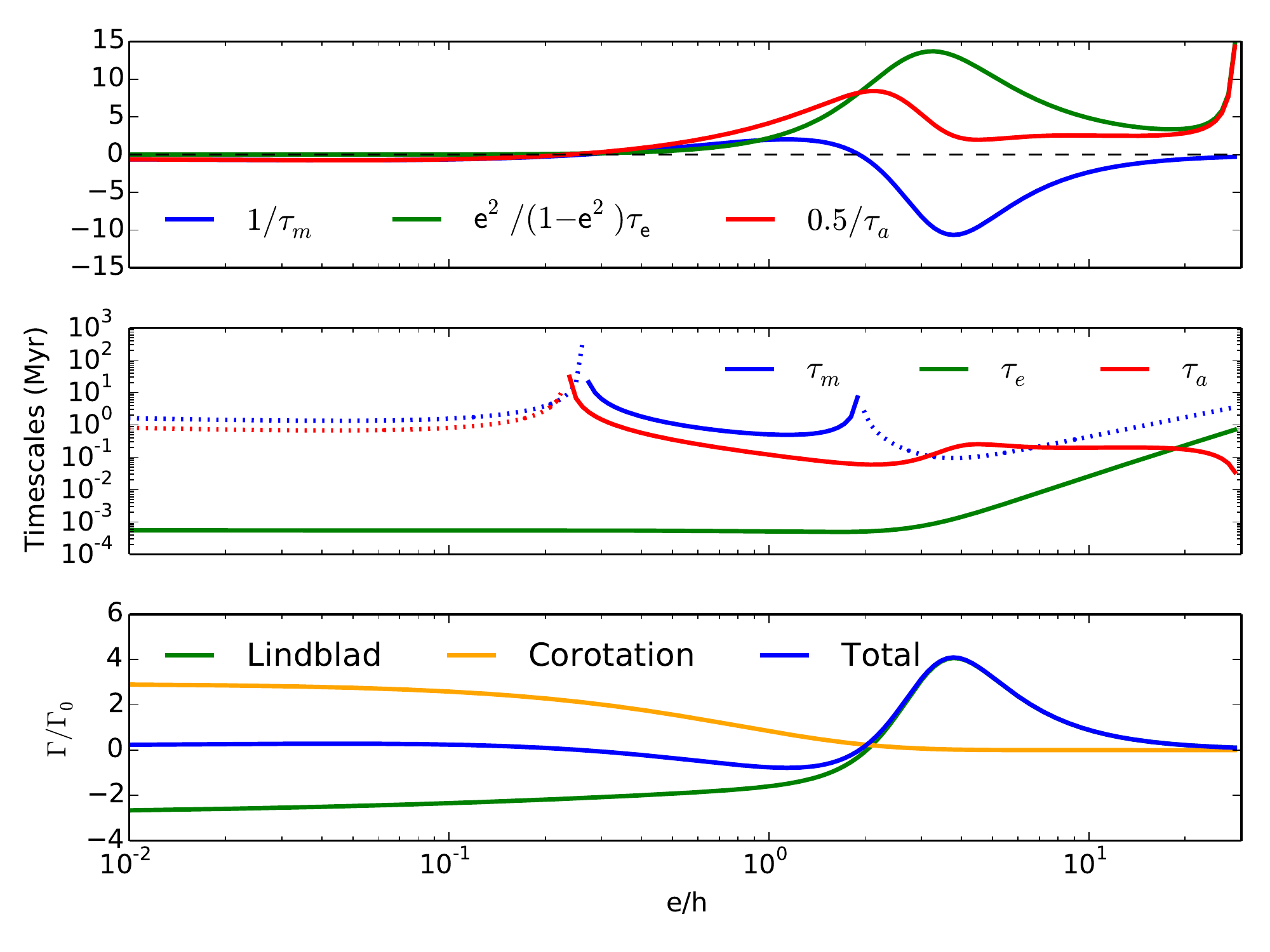}
\end{minipage}\hfill
\begin{minipage}{0.5\textwidth}
\centering
\includegraphics[width=0.99\textwidth]{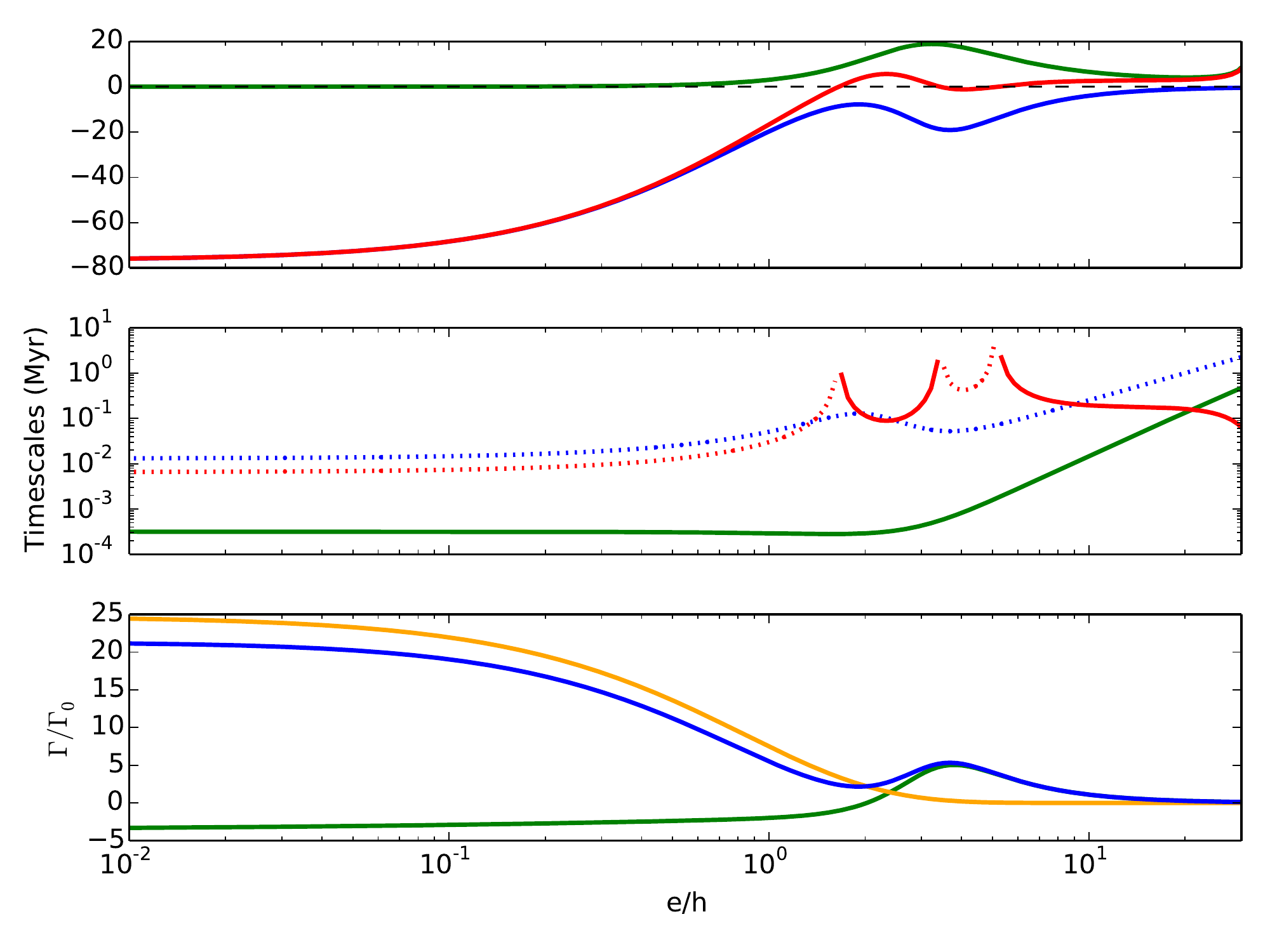}
\end{minipage}
\caption{Plot of various timescales and normalised torques vs eccentricity for a  $1 M_\oplus$ planet. Left: at 1 AU. 
Right: at 0.099 AU. Dotted lines are for negative values.}
\label{fig:migtimes}
\end{figure*}

\subsection{Planet migration near the disc edge}
The reciprocal migration timescale, $0.5\tau_a^{-1}$, and its two components, $\tau_m^{-1}$ and 
$e^2(1-e^2)^{-1}\tau_e^{-1}$, are plotted in the top panels of Fig.~\ref{fig:migtimes} for the fiducial case (all 
supersonic corrections enabled) as a function of planetary eccentricity. The middle panels show the individual 
timescales and the bottom panels depict the normalised torques. The left column pertains to a 1~$M_\oplus$ 
planet at 1~AU while the to the right the planet is at 0.099~AU. The outcome for the linear edge prescription is very 
similar.\\

At 1~AU when $e \ll h$ we generally have $\tau_e \ll \tau_a \sim \tau_m$ and angular momentum transfer between the disc 
and the planet usually leads to {\it inward} migration ($\tau_a>0$); the example in Fig.~\ref{fig:migtimes}, however, 
has outward migration when $e \lesssim 0.2h$ because for this mass and $\alpha$ the strength of the corotation 
torque slightly exceeds that of the Lindblad torque \citep{P11,BBM17}; see bottom of Fig.~\ref{fig:migtimes}. When $e 
\gg h$ we have $\tau_e \sim 2e^2(1-e^2)^{-1}\tau_a \ll \vert \tau_m \vert$ and the {\it inward} migration is the result 
of eccentricity damping at constant angular momentum. These two cases are common to all the combinations of the torque 
formulae. However, the relations for $\tau_e$, $\tau_a$ and $\tau_m$ are different when $e \sim h$ for the various 
configurations. \\

At 0.099 AU (near the disc's inner edge), these two cases are also applicable, except that the difference between 
$\tau_e$ and $\tau_a$ when $e \ll h$ is smaller than at 1 AU.  When $e \sim h$ at 0.099 AU, however, the relations 
for $\tau_e$, $\tau_a$ and $\tau_m$ are different from that at 1 AU.\\

These fundamental differences are clearly seen between the two panels, assuming that equation (\ref{eq:torque}) 
and the supersonic corrections are applicable at the disc edge. \\

Suppose there is a single planet at $r_{\rm tr}$. When a second planet approaches the first one, it is likely to become 
trapped in resonance \citep{P13}. The equilibrium eccentricity of both planets is the result of the balance of 
migration and damping, and is $e_{\rm eq} \approx 1.3 h$ \citep{GS14}. At this eccentricity $\tau_m < 0$, but $\tau_a > 
0$ when $r_{\rm in} < r < r_{\rm tr}$, and the planets migrate {\it inwards}: the innermost planet has its eccentricity 
excited as the outer planet pushes the pair starward until the latter reaches $r_{\rm tr}$; the inner planet is 
now parked inside the disc cavity. Should a third, fourth and additional planets approach the inner pair the mechanism 
will likely repeat itself until the outermost planet is at the disc edge and all the other planets are inside of it, or 
until the outermost planets can no longer push the inner chain deeper into the cavity due to the chain's inertia.\\

The above arguments assume that the planet is surrounded by a gas disk on both sides. Near the edge, 
the torque is one-sided: at the inner disk edge there is only the outer Lindblad torque that pushes the planet inwards, 
but no inner Lindblad torque. \cite{Liu17} implemented simplified one-sided corotation and Lindblad torques and 
showed that small planets could be trapped near the edge, but their implementation only works for an infinitely sharp 
edge.

\section{Numerical methods}
\label{numerics}
To study the behaviour of planets near the disc edge we performed a set of numerical N-body simulations consisting of a 
solar-mass star and four equal-mass planets. These integrations used the symplectic $N$-body code SyMBA \citep{dll98}, 
which was heavily modified to include the effects of eccentricity and inclination damping as well as planet migration 
by the gas disc according to the formulation above \citep{Mat17}.\\

We initially place the planets beyond their 2:1 mean-motion resonances. The planet's masses are all either 0.1, 0.5, 
2, 3, 5 or 8 Earth masses ($M_\oplus$). We compute the torques at each time step for each body; we apply eccentricity 
damping only when $e>0.001h$. There is a surface density maximum at $r_{\rm tr}=0.1$~AU from the star. Closer to the 
star than the disc edge at $r_{\rm in}$ there is no migration nor damping.\\

Simulations are run for 2~Myr with a time step of 0.146~d. Bodies are removed when they are closer than 0.02~AU or 
farther than 100~AU from the star, or when they collide. We assume perfect accretion during collisions. Initially 
$\dot{M}_{*8} = 1$ and $\alpha_3=1$. For simplicity we keep $\alpha$ fixed despite its potential to change close 
to the star where MRI effects \citep{BS13} and disc ionisation are important \citep{Gam96}. The value of $\alpha$ 
affects the torque in a complicated manner as described in equation (\ref{eq:entropypart}) \citep{P11,BBM17}. As we show 
below, the most important factor in altering the torque is the supersonic corrections, in particular the corotation 
reduction, $F_C$, which is independent of $\dot{M}_*$ and $\alpha$. \\

We test the dependence of the migration on eccentricity by selectively enabling or disabling the supersonic 
corrections to the corotation torque, the Lindblad torque and the eccentricity damping timescale. Apart from 
\cite{Masset06} we are not aware of any systemic hydrodynamical studies of embedded planets near the disc edge. 
Motivated by this, we enable or disable ('on' or 'off') the corotation reduction term ($F_C$ in equation 
(\ref{eq:fcfl}), 'CR') as well as the supersonic corrections to the eccentricity damping timescale (the term inside the 
parentheses in equation (\ref{eq:taue}), 'ED') and to the Lindblad torque ($F_L$ in equation (\ref{eq:fcfl}), 
'EL'). This simple approach should suffice in the absence of more detailed torque prescriptions near the edge.

\section{Results}
\begin{figure*}[t]
\resizebox{\hsize}{!}{\includegraphics[]{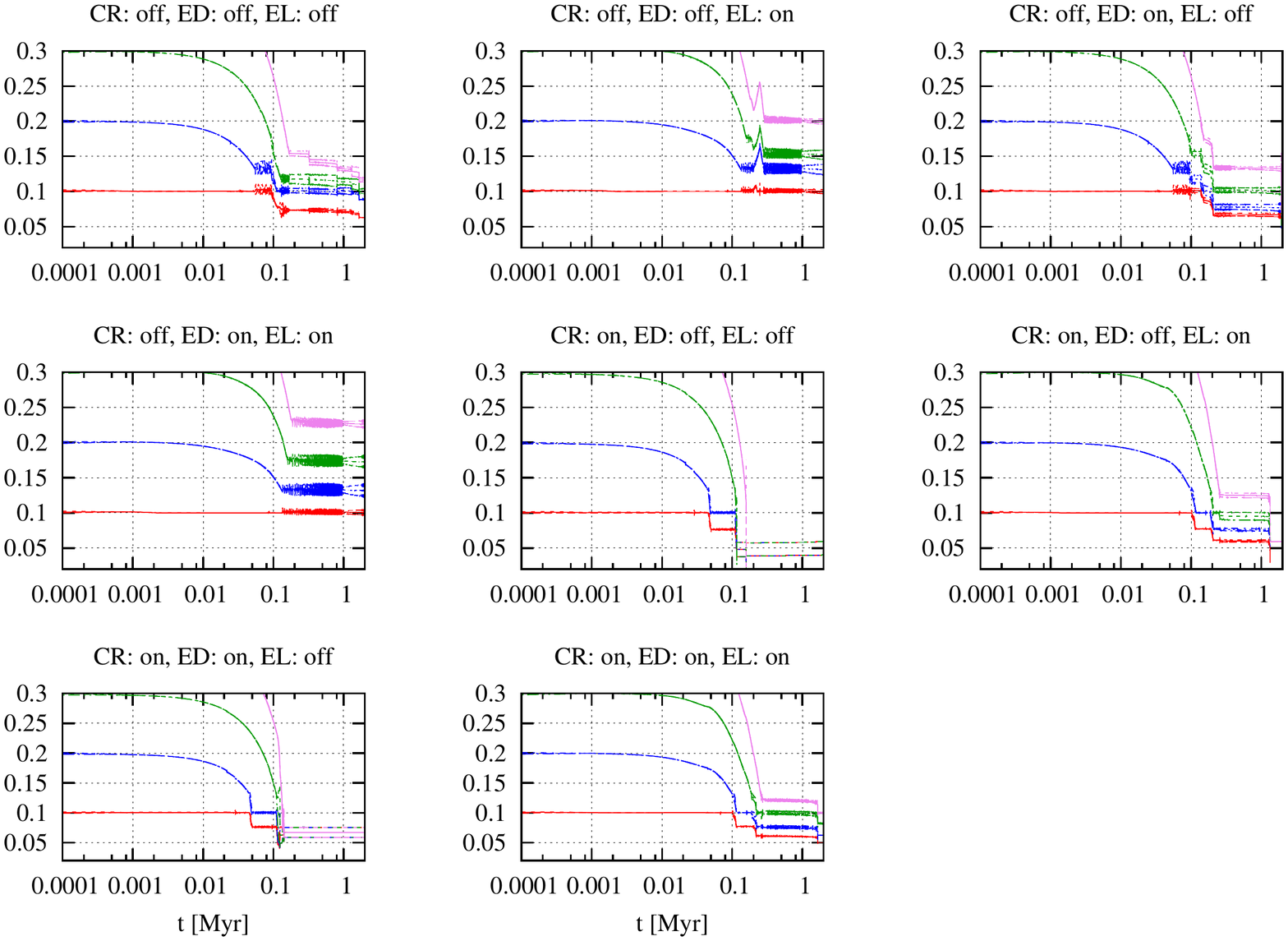}}
\caption{Migration of four 2 $M_\oplus$ planets towards the disc edge. The panels all depict different migration 
prescriptions.}
\label{fig:evo4-2me}
\end{figure*}

We show an example of the evolution of four planets of $2M_\oplus$ near the disc edge in Fig.~\ref{fig:evo4-2me}. The 
evolution is very different depending on the migration prescription. In the broadest sense the planets become trapped 
in resonances outside the disc edge when the corotation reduction $F_C$ is {\it not} applied, and preferably when the 
supersonic correction to the Lindblad torque, $F_L$, {\it is} applied. If either or both of these are applied, multiple 
planets will end up inside the disc cavity.\\

This outcome is in contrast with \cite{I17} because they use $\tau_m$ to compute the semi-major 
axis evolution of the planets instead of $\tau_a$. Since $\tau_m < 0$ at the disc edge for any value of $e$, their 
innermost planet can often stall any additional incoming planets even if it or the others are supersonic, though 
the exact evolution is mass dependent. For example, in their Figure 4, the innermost planet is 10 $M_\oplus$ and all 
planets are trapped beyond 0.1~AU, while in their Figure 5 the innermost planet (initially) has a few $M_\oplus$ and a 
chain of planets is pushed inward as more massive planets migrate in. In resonance the approximation $\tau_a = 
\tau_m$ breaks down because the planets are supersonic; equation (\ref{eq:ta}) should be used instead.\\ 

Our results are also in disagreement with \cite{O10}, who found that a sharp edge was able to prevent the planets from 
falling into the cavity. \cite{O10} concluded that the imbalance of increased drag on the planet at aphelion versus 
little to no drag at perihelion caused the planet to be stationary at the edge with a non-zero eccentricity. However, 
they did not include any supersonic corrections to their migration formulae. \\

In our approach the edge is fairly sharp, but the the tanh function quickly flattens beyond $\tanh 
1\approx 0.76$. Thus if the planet is near $r_{\rm tr}$ the drag at aphelion and perihelion is within 25\% if 
the eccentricity is $e \sim h$. Only when the planet is roughly halfway between $r_{\rm in}$ and $r_{\rm tr}$ is the 
drag at aphelion (where the tanh function is $\sim$1) much stronger than at perihelion (where it is $\sim$ 0) and do we 
possibly recover the situation from \cite{O10}, but only if the supersonic reduction of the corotation torque is 
ignored.\\

\begin{figure*}[t]
\centering
\begin{minipage}{0.33\textwidth}
\centering
\includegraphics[width=0.99\textwidth]{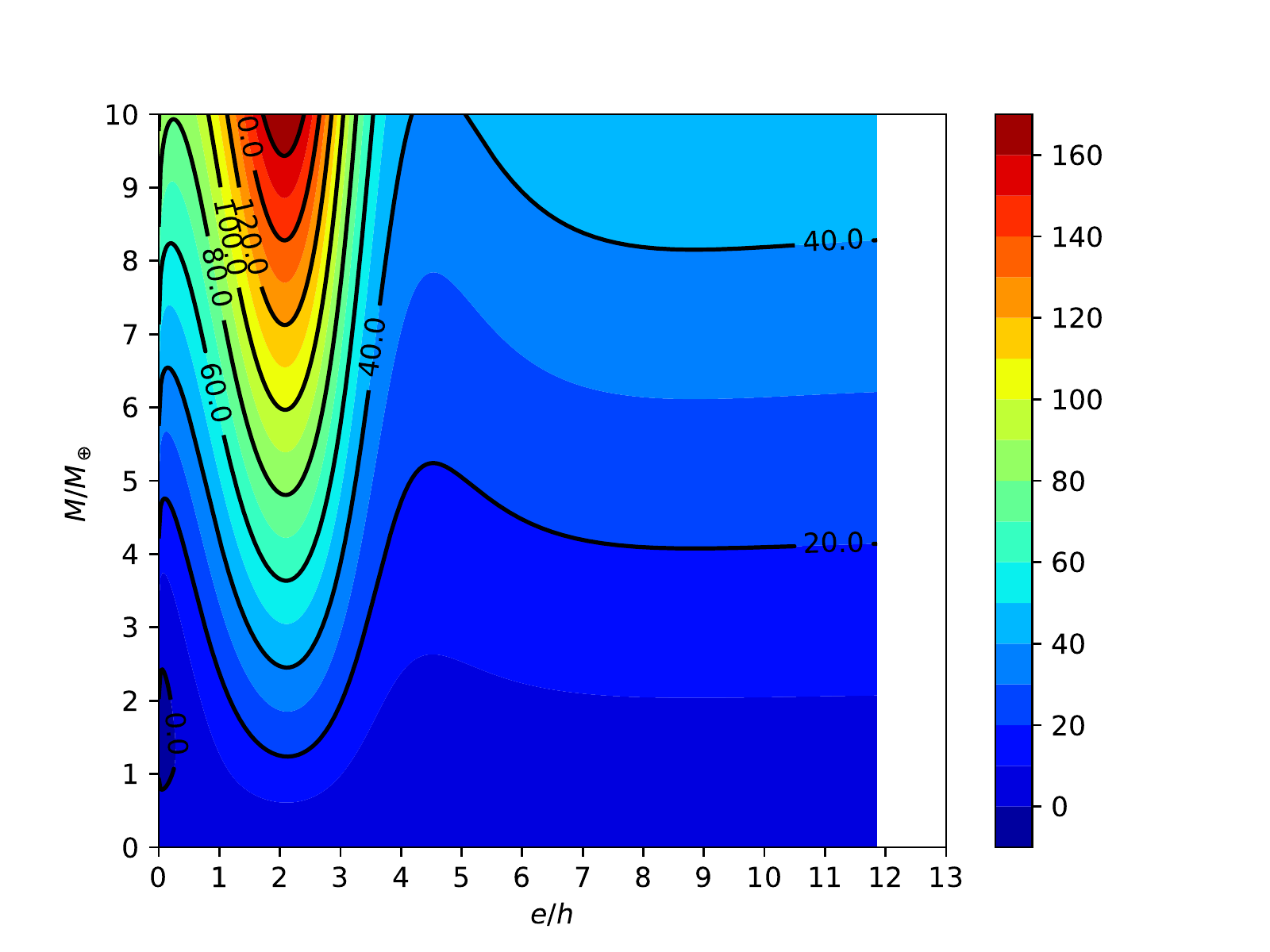}
\end{minipage}\hfill
\begin{minipage}{0.33\textwidth}
\centering
\includegraphics[width=0.99\textwidth]{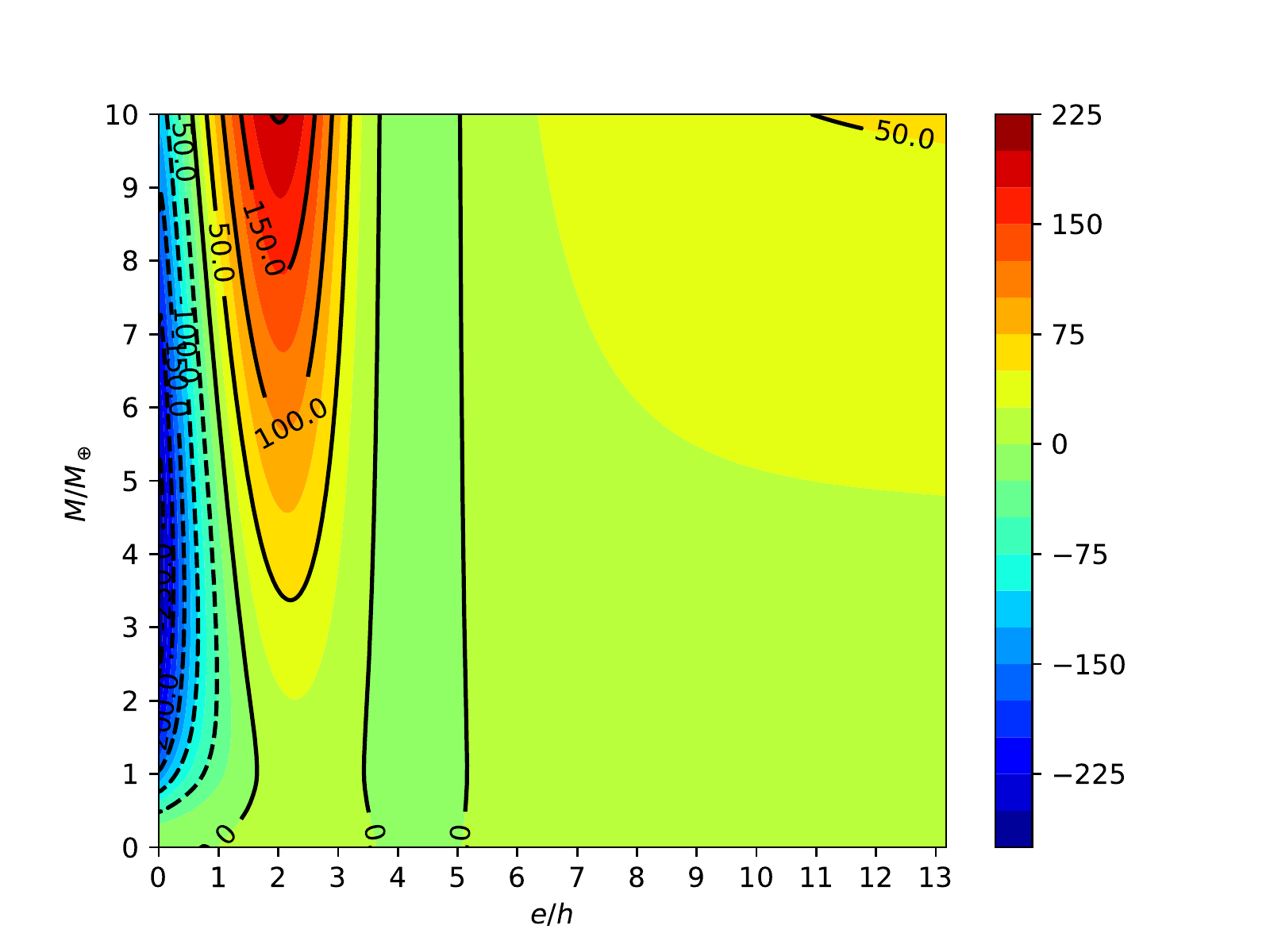}
\end{minipage}\hfill
\begin{minipage}{0.33\textwidth}
\centering
\includegraphics[width=0.99\textwidth]{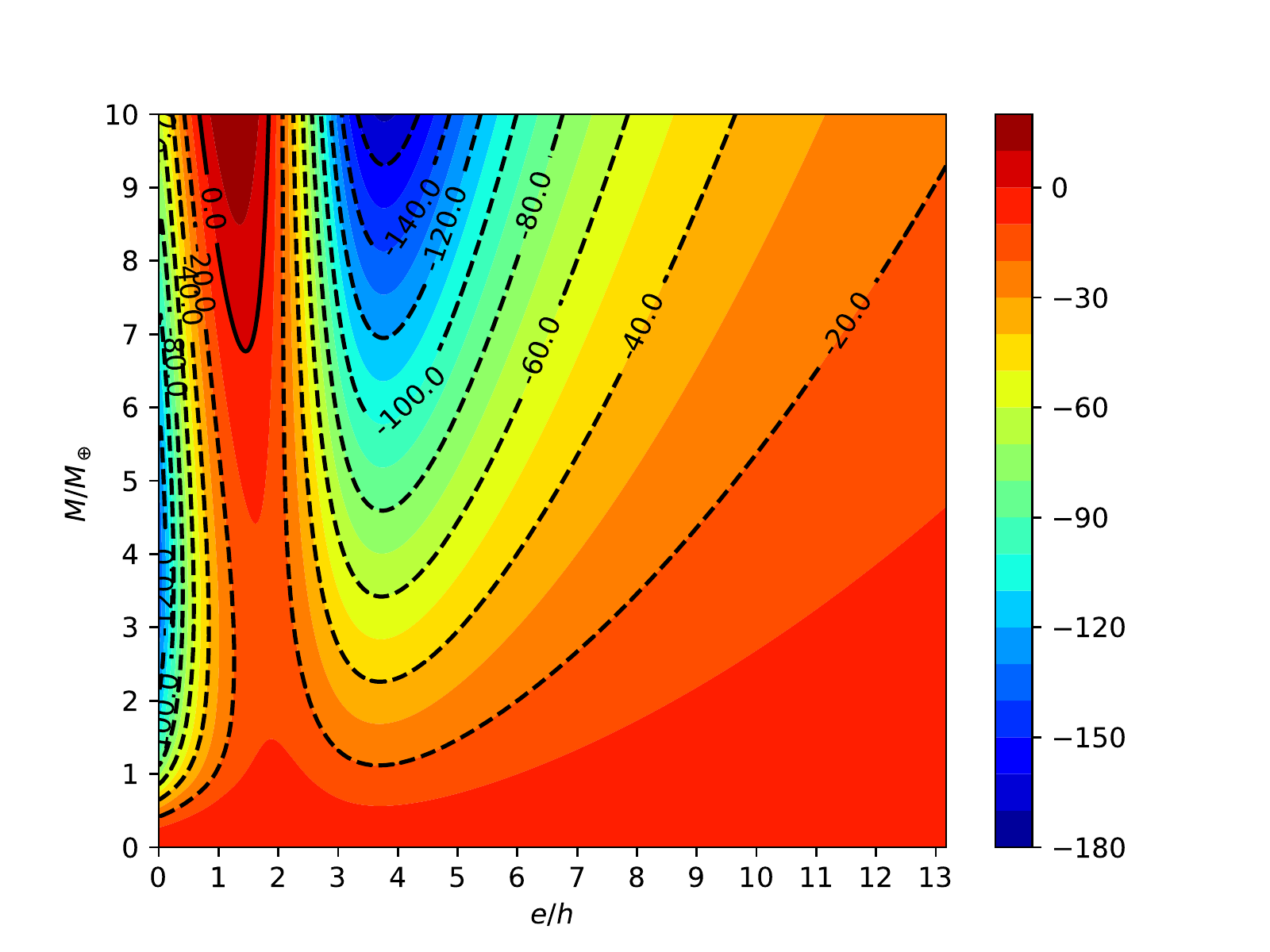}
\end{minipage}\\
\begin{minipage}{0.33\textwidth}
\centering
\includegraphics[width=0.99\textwidth]{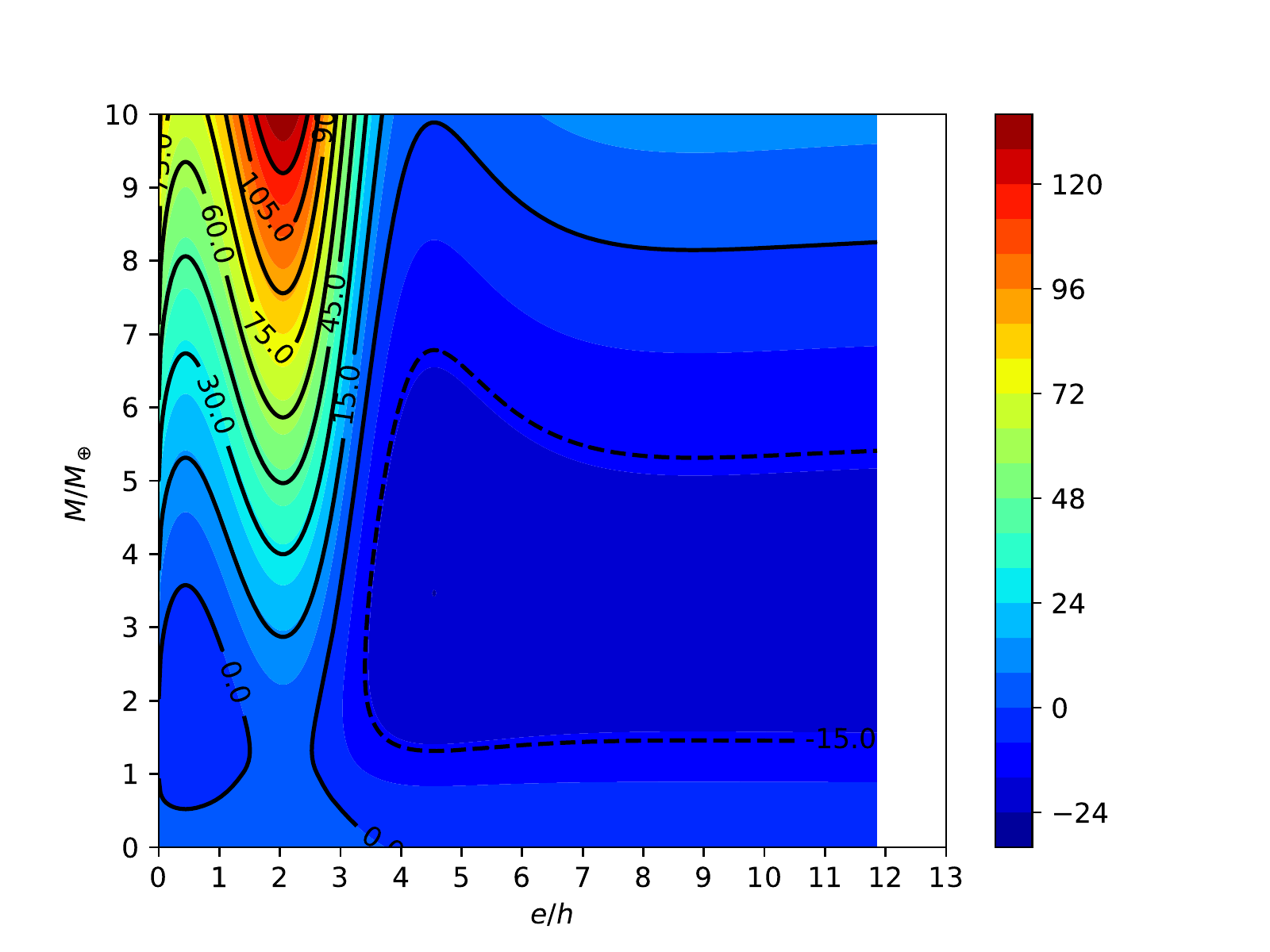}
\end{minipage}\hfill
\begin{minipage}{0.33\textwidth}
\centering
\includegraphics[width=0.99\textwidth]{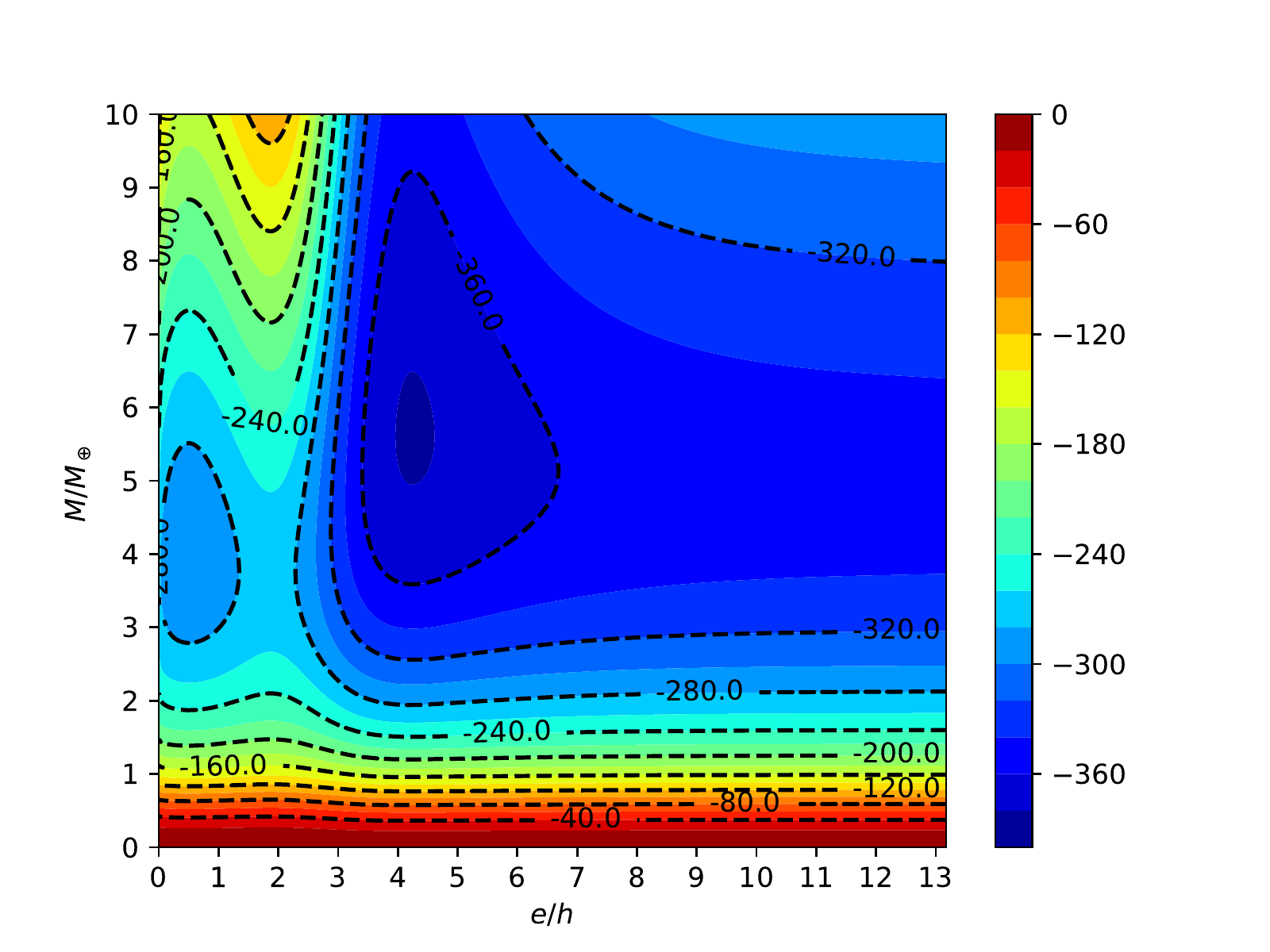}
\end{minipage}\hfill
\begin{minipage}{0.33\textwidth}
\centering
\includegraphics[width=0.99\textwidth]{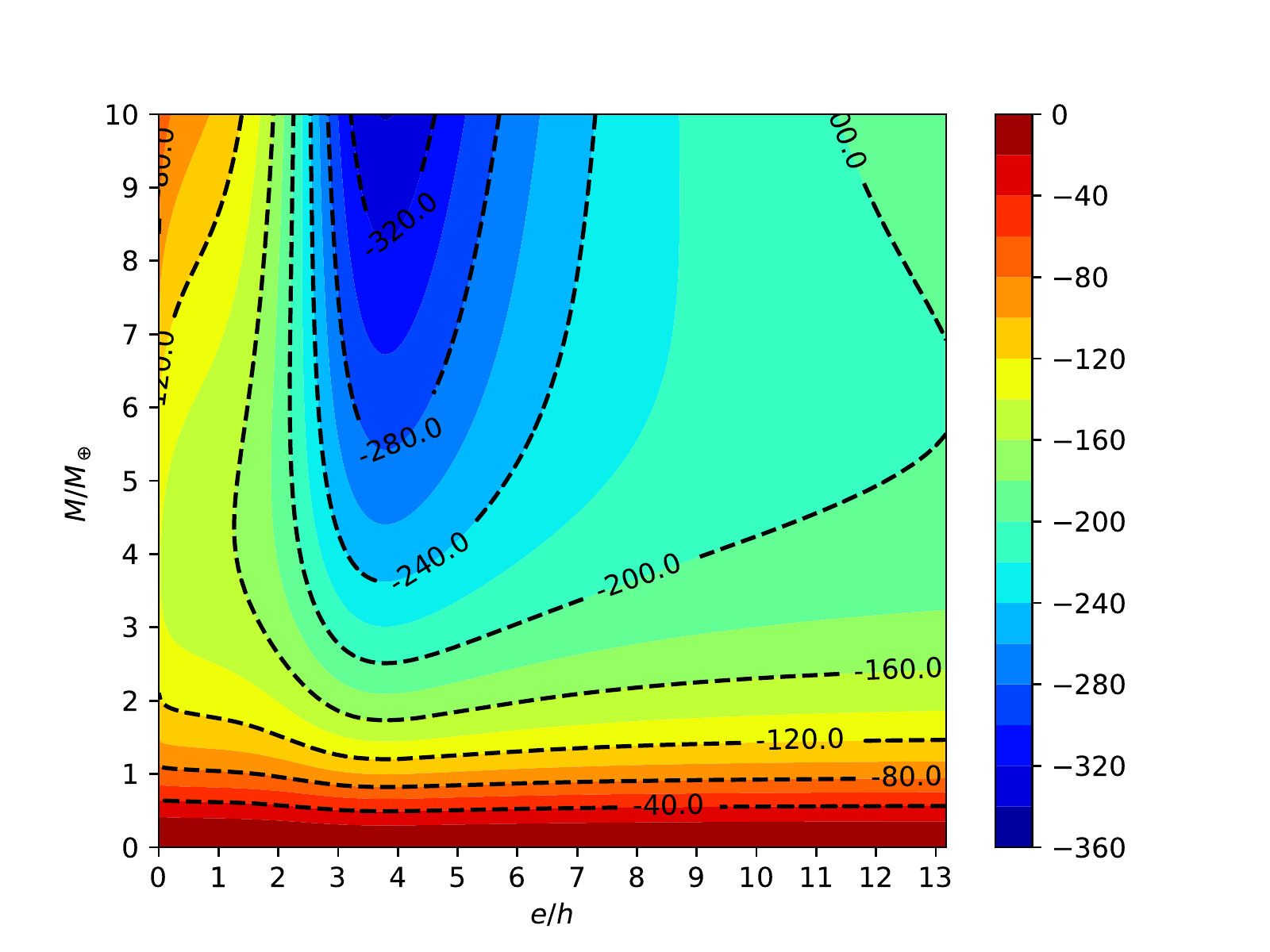}
\end{minipage}
\caption{Left: Contour map of $\tau_a^{-1}$ in Myr$^{-1}$ for a planet of 1$M_\oplus$ at 1 AU as a function of 
$e/h$ and $m_p$. Middle: same at 0.099 AU. Right: $\tau_m^{-1}$ in Myr$^{-1}$ at 0.099 AU. Bottom panels are the same 
but now $F_C=1$ i.e. there is no corotation reduction at high eccentricity.}
\label{fig:tamap}
\end{figure*}

\subsection{Different torque prescriptions yield different outcomes}
What aspects of the torque prescription are responsible for the different behaviour at the edge? At low eccentricity 
outward migration is caused by the corotation torque while inward migration is caused by the Lindblad torque; the 
latter only depends on the temperature and surface density gradients \citep{P11}. When supersonic corrections are 
considered the total torque also depends on $e$. In general with our prescription $\tau_a^{-1}$ is not a 
monotonically decreasing function of $e$, but instead has a peak near $e \sim h$ and a trough at $e \sim 4h$ 
\citep{cn08}. This behaviour is caused by how $\tau_m$ varies with eccentricity, and also because the maxima of 
$\tau_m^{-1}$ and $e^2/(1-e^2)\tau_e^{-1}$ do not coincide (top-right panel of Fig.~\ref{fig:migtimes}). The migration 
rate peaks for all planetary masses near $e \sim 0.5h$, reaches a minimum when $e \sim 2h$ and a further maximum when $e 
\sim 4.5h$. This non-monotonic behaviour is inconsistent with the analytical results of \citep{Muto11}.\\

However, the behaviour is different when the planet is inside of the trap. At very low eccentricity, and 
for nearly all planet masses, $\tau_a^{-1}$ is negative and low, implying slow outward migration. Migration is 
inward when $e \sim h$ and outward again at higher eccentricity. The top row of Fig.~\ref{fig:tamap} is a contour map 
showing $\tau_a^{-1}$ at 1~AU (left), near the disc's inner edge (middle), together with $\tau_m^{-1}$ near the 
edge (right).\\

The region of inward migration near moderate $e$ prevents the trap from stalling migrating planets in resonances, and 
any planet situated at $r_{\rm tr}$ is pushed deep inside the cavity, together with any planets interior to it. 
Therefore, the outcome of numerical simulations and the ability to trap planets near the disc edge depends on the exact 
migration prescription employed. When using $\tau_m$ the migration is always outwards.\\ 

Figure~\ref{fig:evo4-2me} suggests that eliminating the corotation reduction, $F_C$, and weakening the Lindblad torque 
by applying $F_L$ provides the best prescription to trap multiple planets in resonance outside of $r_{\rm tr}$, 
assuming the current torque formulae hold near the edge (cf. \cite{Liu17}). The bottom row of Fig.~\ref{fig:tamap} 
shows similar contour maps but now $F_C=1$ i.e. there is no corotation reduction. The behaviour is qualitatively 
different everywhere: at the edge migration is always outward, but the strength is a complicated function of both the 
planetary mass and the eccentricity.\\

Hydrodynamical simulations show that the corotation torque weakens as the eccentricity increases and mostly disappears 
at $e \gtrsim 3h$ \citep{CN07}. It thus appears to be unphysical to remove the corotation reduction far from the disc 
edge, but it is unclear if this removal is applicable near the edge. The exponential reducion of \cite{FN14} does not 
appear to hold for low values of $h \lesssim 0.05$; the corotation reduction also depends on how the torque is measured. 
Their Figs. 4 and 9 clearly show torque maxima near $e \sim 2.5h$ so that the exponential reduction may not be 
universally applicable. Clearly more work is needed, both on the reduction itself but also how it behaves near the disc 
edge.\\

\begin{figure*}[t]
\resizebox{\hsize}{!}{\includegraphics[]{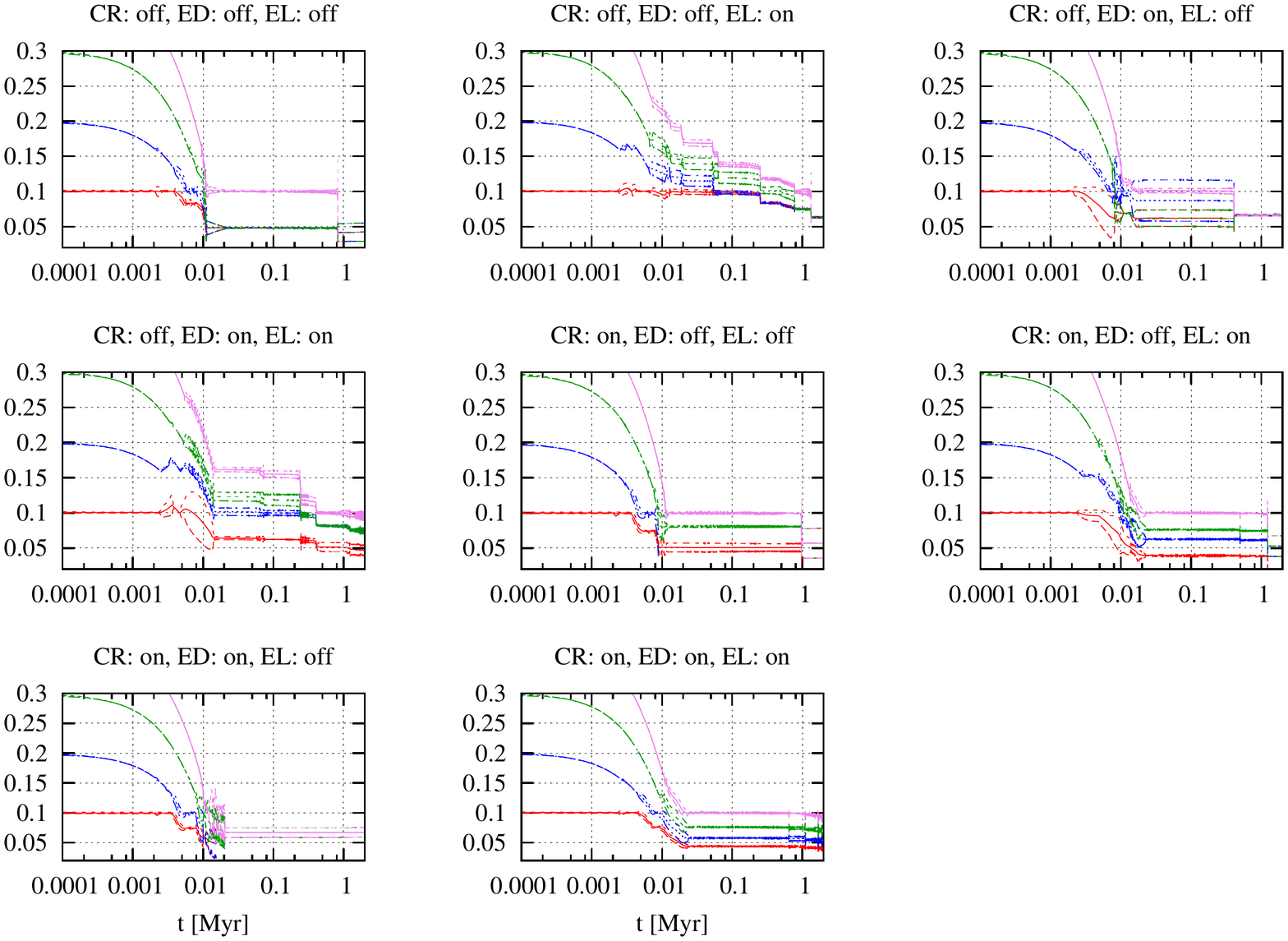}}
\caption{Same as Fig.~\ref{fig:evo4-2me} but the planets are now 5$M_\oplus$.}
\label{fig:evo4-5me}
\end{figure*}

In Fig.~\ref{fig:evo4-2me} the bottom-middle panel with all the supersonic effects enabled was able to temporarily trap 
the planets in a resonance even though the innermost planets were pushed into the cavity. Increasing the corotation 
torque kept the planets outside of the cavity (CR: OFF). This structure does not hold for higher-mass planets 
because they migrate faster and therefore excite themselves to higher eccentricities once the innermost planet is in 
the cavity and eccentricity damping is weak or non-existent. An example is shown in Fig.~\ref{fig:evo4-5me}, which is 
the same as Fig.~\ref{fig:evo4-2me} but now the planets are 5$M_\oplus$. 

\section{Conclusions}
The ability to trap multiple low-mass planets in a resonant chain outside the inner edge of the protostellar disc has 
been investigated. These low-mass planets execute type I migration which pulls them invariably towards the star. In 
the absence of a barrier these would all collide with the star. The disc's inner edge could provide a trapping 
mechanism \citep{Masset06}. We have tested two types of sharp inner edges of the disc: a hyperbolic tangent and a linear 
function, along with different migration prescriptions.\\

We find that a neat pileup of resonant planets outside the disc edge is established if the corotation torque does 
not rapidly diminish at high eccentricity. The expectation is that if the resonant chain of the planets remains outside 
the inner disc edge they eventually start orbit crossing and instigate a phase of giant impacts. This may account 
for formation of similar-sized, regularly spaced, non-resonant low-mass planets that are found to be common in 
relatively close-in regions by Kepler observations. However, the eccentricity damping and disc torques in the 
supersonic regime remain uncertain near the disc’s inner edge. Due to resonant interactions, eccentricity is generally 
excited to values $e \sim h$ for which the migration is generally inward. Therefore we call for detailed studies on 
eccentricity damping and disc torques in the supersonic regime and near the disc edge. Such a study will play an 
important role in understanding the common architecture of compact systems.

\end{document}